\documentclass[12pt]{article}

\usepackage{amsmath}
\usepackage{hyperref}
\usepackage{mathrsfs}

\begin{document}

\title{Are the equations of motion more fundamental than the conservation of energy in mechanics?}
\author{        V. Tanr{\i}verdi \\
tanriverdivedat@googlemail.com 
}

\date{\today}

\maketitle

\begin{abstract}

In some cases, it is possible to show the conservation of energy by using equations of motion in mechanics.
By considering these results, some people can think that the conservation of energy is the result of equations of motion or Newton's second law.

If we consider the conservation of energy by itself, it is valid for nearly all natural sciences and more general than equations of motion.
From this perspective, it is not totally convenient to say that the conservation of energy is the result of equations of motion.

It is clear that there can be a relation between them, but it is not explicit enough.
In this study, we have studied the relationship between the conservation of energy and equations of motion.
And, the study revealed a subtle difference between them in mechanics which can be used to experimentally test which explains the nature best or to better understand the relation between them. 

\end{abstract}

\section{Introduction}

The fundamental laws constitute the basis of scientific understanding.
The conservation of energy and Newton's second law or equations of motion, which can be obtained from Lagrange equations, can be considered as two of the fundamental laws of classical mechanics.
In general, these two laws are considered separate fundamental laws.
This can be seen as natural, but there are some works considering their relations.

The history related to these laws is a long one and some parts are related to the "vis viva controversy" including Newton's second law and Leibniz's usage of mass times velocity square \cite{Papineau,Iltis,Terrall}.
In the beginning, there were different camps; some of the scientists were following Leibniz's thoughts, some of them were following Newton's and some others were defending other thoughts.
As time passed, the situation changed.
Most scientists had started to use Newton's approach since Newton's method was successful and Leibniz's formulation had some problems.
Later, the conservation of energy for mechanics is shown to be true by the work of scientists like Huygens, Young, Coriolis, Mayer, Joule, Helmholtz, W. Thomson (Lord Kelvin) and many others \cite{Coriolis, Helmholtz, Elkana, Mach}.
There were also other scientists trying to understand the relationship between Newton's second law and the conservation of energy.
D'Alembert, Lagrange and Carnot were aware that the conservation of energy can be seen as a consequence of Newton's second law in some special cases \cite{Ardema}.
These scientists took Newton's second law as the basis and did not consider the other way around.
This is, probably, because of learning Newton's second law prior to the conservation of energy and trying to understand it in terms of Newton's second law.

Within centuries, we have learned lots of things, and Lagrangian formalism or equations of motion is used successfully in nearly all subfields of physics. 
If we consider this and see that it is possible to obtain the conservation of energy by using Newton's second law, we may regard it as more fundamental than the conservation of energy.
But, there are other works related to the topic which should be taken into account. 

One of these is Hamilton's work related to obtaining an equivalent set of equations to equations of motion.
Hamilton developed a formalism to obtain equations describing the system, Hamilton equations, from Hamiltonian which is equal to energy "if the equations defining the generalized coordinates don't depend on time explicitly", "and if the forces are derivable from a conservative potential" \cite{Goldstein}.  
Hamilton equations will provide equation sets for each momentum. 
If we consider the cases where Hamiltonian is equal to energy, we have a formalism giving equations describing the system from energy which we will reconsider at the conclusion.

There are also some recent studies related to this topic.
Vinokurov gives a formalism, the principle of detailed energy conservation, to obtain equations of motion from a Lagrangian which is obtained in terms of energy by using the conservation of energy in his work \cite{Vinokurov}.
Carlson writes energy in terms of generalized coordinates and momenta, and then gives a formalism to get equations of motion by using the conservation of energy, and states that his formulation does not describe properly electromagnetic field and relativistic cases \cite{Carlson}.
Lindgren presents a prescription, generalized energy method, to obtain equations of motion which is formed by using the relation that the work done by a force is equal to change in the kinetic energy, which can be considered as the basis of the conservation of energy, and claims that equations of motion can be obtained except the cases including Coriolis force and gyro moments \cite{Lindgren}.
Zhou and Wang claim that they obtain equations of motion in different cases by using the conservation of energy \cite{ZhouWang}.
Hanc and Taylor give a summary of Newtonian physics including Lagrange equations and the conservation of energy, and obtain the equation of motion for a one-dimensional motion from the conservation of energy while giving this summary \cite{HancTaylor}.
On the other hand, Neuenschwander, Taylor and Tujela state that variation of energy is not a suitable quantity to obtain equations of motion \cite{NTT}.

As it can be seen there are various approaches to the problem: Some scientists obtain the conservation of energy by using equations of motion, some scientists are trying to obtain equations of motion from the conservation of energy and some of them argue against it.
In this work, we will study the relationship between the conservation of energy and equations of motion and try to clarify the topic.
Though some authors claim that equations of motion can be obtained from the conservation of energy, we show that it is possible only in some cases.
More importantly, the results show that there is a subtle difference between the conservation of energy and Newton's second law or equations of motion.
And, this difference is worth to further study and can help to elucidate the situation.

\section{Theory}
\label{derivation}

One can get equations of motion from Lagrange's equations which are given by \cite{Goldstein}
\begin{equation}
        \frac{d}{d t}\left( \frac {\partial \mathscr{L}}{\partial \dot q_i} \right)-\frac{\partial \mathscr{L}}{\partial q_i}=0,
\end{equation}
where $\mathscr{L}=T-U$ is Lagrangian, $T$ is the kinetic energy, $U$ is the potential energy, $t$ is the time, $q_i$ and $\dot q_i$ are generalized coordinates and velocities, respectively.

Now let us consider energy, $E(q_i, \dot q_i)=T+U$.
For energy-conserved systems, its total time derivative is equal to zero.
In this case, one can write
\begin{equation}
        \label{dedt3d}
        \sum_i \left[\frac {\partial E}{\partial \dot q_i}\frac {d \dot q_i}{d t}+ \frac {\partial E}{\partial q_i}\frac {d q_i}{d t} \right]=0.
\end{equation}
This equation is one of the starting points of previous works that are trying to relate the conservation of energy with equations of motion \cite{Vinokurov, Carlson}.
It is hard to compare this equation with Lagrange equations in these forms.
Forasmuch as we will consider scleronomic systems having velocity-independent potential energy, $U=U(q_i)$.
For scleronomic systems with many particles, the kinetic energy can be written as $T=\frac{1}{2} \sum_{jk} M_{jk} \dot q_j \dot q_k $ where $M_{jk}=\sum_i m_i \frac{\partial \vec r_i}{\partial q_j} \cdot \frac{\partial \vec r_i}{\partial q_k}$, where $m_i$ and $\vec r_i$ are the mass and position vector of $i$th particle, respectively \cite{Goldstein}.
Then, Lagrange equations for $i$th particle can be found as
\begin{equation}
        \sum_k M_{ik} \ddot q_k +\frac{1}{2} \sum_{jk} \frac{\partial M_{jk}}{\partial q_i} \dot q_j \dot q_k+ \frac{\partial U}{\partial q_i}=0,
\end{equation}
and from Eq. \eqref{dedt3d}, one can obtain
\begin{equation}
        \sum_i \dot q_i \left[\sum_k M_{ik} \ddot q_k +\frac{1}{2} \sum_{jk} \frac{\partial M_{jk}}{\partial q_i} \dot q_j \dot q_k+ \frac{\partial U}{\partial q_i}\right]=0. \label{dedtez}
\end{equation}
This equation, obtained from the total time derivative of energy, has a summation over index $i$ and this is its main difference from equations of motion.
Differently from equations of motion, the summation of all terms can be equal to zero. 

From these two equations, similar to D'Alembert, Lagrange and Carnot, one may conclude that Lagrange equations are enough to show the conservation of energy for considered systems since the generalized velocity $\dot q_i$ is different from zero in general.
This is also shown previously for some cases without considering scleronomic systems \cite{MarionThornton, Arnold}.
But, the mentioned conclusion is the result of approaching the problem from only one perspective.
For a full scientific inquiry, this topic requires further consideration and it is possible to ask: Is the result of conservation energy different from equations of motion?

Now, we will try to find an answer to this question.
If the motion is one-dimensional, then the summation over index $i$ in Eq. \eqref{dedtez} drops.
And, since $\dot q_i \ne 0$ in general, one obtains the same equation as Lagrange equations.
Similar considerations are also valid for two or three-dimensional systems where by using conserved momenta, one can write Lagrangian and energy in a one-dimensional form.
One can easily see these by studying central force motion and the heavy symmetric top.

Multi-dimensional cases, where the reduction is impossible, give more interesting results.
As an example, one can consider the double pendulum whose kinetic and potential energies can be written as $T=\frac{1}{2}(m_1+m_2)l_1^2 \dot \phi_1^2+\frac{1}{2}m_2 l_2^2 \dot \phi_2^2+m_2 l_1 l_2 \dot \phi_1 \dot \phi_2 \cos (\Delta \phi)$ and $U=(m_1+m_2)g l_1 (1-\cos \phi_1)+m_2 g l_2 (1- \cos \phi_2)$, respectively, where $m_i$ are masses of bobs, $l_i$ are lengths of pendula, $\phi_i$ are deviations from the equilibrium position and $\Delta \phi=\phi_1- \phi_2$ \cite{Taylor}.
By using Eq. \eqref{dedt3d}, one can obtain
\small
\begin{eqnarray}
	&\dot \phi_1 & \left[ \frac{m_1+m_2}{m_2} \frac{l_1}{l_2} \ddot \phi_1 + \ddot \phi_2 \cos(\Delta \phi) + \frac{m_1+m_2}{m_2} \frac{g}{l_2} \sin \phi_1 + \dot \phi_2^2 \sin(\Delta \phi) \right]  \nonumber \\
	&+ &\dot \phi_2  \left[ \frac{l_2}{l_1} \ddot \phi_2 + \ddot \phi_1 \cos(\Delta \phi) + \frac{g}{l_1} \sin \phi_2 - \dot \phi_1^2 \sin(\Delta \phi)\right]  =0 \label{ecdp} 
\end{eqnarray}
\normalsize
From Lagrange equations, equations of motion can be obtained as \cite{Shinbrot}
\small
\begin{eqnarray}
	& &\frac{m_1+m_2}{m_2} \frac{l_1}{l_2} \ddot \phi_1 + \ddot \phi_2 \cos(\Delta \phi) + \frac{m_1+m_2}{m_2} \frac{g}{l_2} \sin \phi_1 + \dot \phi_2^2 \sin(\Delta \phi)=0 \label{eom1dp}\\
	& &\frac{l_2}{l_1} \ddot \phi_2 + \ddot \phi_1 \cos(\Delta \phi) + \frac{g}{l_1} \sin \phi_2 - \dot \phi_1^2 \sin(\Delta \phi) =0. \label{eom2dp} 
\end{eqnarray}
\normalsize
Here, Eq. \eqref{ecdp} and Lagrange equations are not equal to each other.
However, it can be seen that the conservation of energy, i.e. $\frac{d E}{d t}=0$, can be obtained from equations of motion by using the fact that it is equal to $\dot \phi_1$ times Eq. \eqref{eom1dp} plus $\dot \phi_2$ times Eq. \eqref{eom2dp}, and according to Lagrange equations Eq.s \eqref{eom1dp} and Eq. \eqref{eom2dp} are equal to zero.
On the other hand, though the conservation of energy can be obtained from equations of motion, the equation that is obtained from the conservation of energy, i.e. Eq. \eqref{ecdp}, is different and does not give equations of motion. 
There is a subtle point here which will be considered at the conclusion.

\section{Conclusion}

We have studied the relation between the conservation of energy and equations of motion for scleronomic cases with velocity-independent potentials.
We have seen that one can get the same equations from the conservation of energy and equations of motion for one-dimensional cases and some multi-dimensional cases where energy can be written in one-dimensional form by using conserved momenta.

For multi-dimensional cases, whose energy cannot be written in one-dimensional form, the resultant equations are not the same.
This difference shows that Lagrange equations and the conservation of energy do not designate the same results, in general.
To see the difference, we have considered the double pendula as an example.

We have mentioned in the introduction that Hamiltonian is equal to the energy in some cases and Hamilton equations, a set of equations for each momenta, can be obtained from it.
If we consider the double pendula example, there will be two sets of equations for each momenta.
On the other hand, from the conservation of energy, we obtain only a single equation.
Then we can easily say that Hamilton equations and the equation obtained from the conservation of energy are different things.

There are some recent studies concerning the double pendula, and only some of these consider the conservation of energy \cite{Vadai, Korsch, Rafat}.
However, none of these studies include any results considering any possible difference between Eq. \eqref{ecdp} and Eq.s \eqref{eom1dp}\& \eqref{eom2dp}.
Nevertheless, by studying experimentally the double pendula or a similar case, one can find an answer to the previously asked question: Is the result of conservation energy different from equations of motion?

An experiment can obey the equation obtained from the conservation of energy, i.e. Eq. \eqref{ecdp}, while not obeying equations of motion, i.e. Eq.s \eqref{eom1dp} and \eqref{eom2dp}. 
If this case holds, then one can say that the conservation of energy is more general, and equations of motion or Newton's second law is a mere consequence of the conservation of energy for cases where either energy is one-dimensional or energy can be written in the one-dimensional form.
A single experimental result is enough to say this.

On the other hand, an experiment can obey equations of motion, i.e. Eq.s \eqref{eom1dp} and \eqref{eom2dp}, and the conservation of energy is satisfied as a natural result of this.
In this case, a single result is not enough to say that the conservation of energy is the result of equations of motion since it does not cover all possibilities.
For elucidation, we require experimental and further theoretical studies.


\begin{thebibliography}{99}

        \bibitem{Papineau}
                Papineau, D. 1977 The vis viva controversy: Do meanings matter? \textit{Studies in History and Philosophy of Science Part A} \textbf{8(2)} pp. 111-142
                DOI 10.1016/0039-3681(77)90011-5

        \bibitem{Iltis}
                Iltis, C. 1971 Leibniz and the Vis Viva Controversy \textit{Isis} \textbf{62} pp. 21-35
                DOI 10.1086/350705

        \bibitem{Terrall}
                Terrall, M. 2004 Vis Viva Revisited \textit{History of Science} \textbf{42} pp.189-209
                DOI 10.1177/007327530404200202

	\bibitem{Coriolis}
		Coriolis, G. G. 1829 \textit{Du Calcul de l'Effet des Machines} (Paris: Carilian-Goeury)

	\bibitem{Helmholtz}
		Helmholtz, H. 1853 On the conservation of force \textit{Scientific Memoirs, Selected from the Transactions of Foreign Academies of Science, and from Foreign Journals. Natural Philosophy} (London: Taylor and Francis) \textbf{1} p. 114

	\bibitem{Elkana}
		Elkana, Y. 1974 \textit{The Discovery of the Conservation of Energy} (Cambridge: Harvard University Press)

	\bibitem{Mach}
		Mach, E. 1894 On the principle of the conservation of energy \textit{The Monist} \textbf{5(1)} pp. 22-54

	\bibitem{Ardema}
		Ardema, M. D. 2005 \textit{Analytical Dynamics: Theory and Applications} (New York: Kluwer Academic/Plenum Publishers) p. 157

        \bibitem{Goldstein}
                Goldstein H., Poole C. and Safko J. 2002 \textit{Classical Mechanics} 3rd Ed (New York: Addison-Wesley) pp. 339, 45, 25, 74, 208-212

        \bibitem{Vinokurov}
                Vinokurov, N. A. 2014 Analytical mechanics and field theory: derivation of equations from energy conservation \textit{Phys. Usp.} \textbf{57(6)} pp 593-596
                DOI 10.3367/UFNe.0184.201406c.0641

        \bibitem{Carlson}
                Carlson, S. 2016 Why not energy conservation? \textit{Eur. J. Phys.} \textbf{37} p. 015801
                DOI 10.1088/0143-0807/37/1/015801

        \bibitem{Lindgren}
                Lindgren E. R. 2002 The Generalized Energy Method for the Formulation of the Equations of Motion in Classical Mechanics \textit{Phys. Scr.} \textbf{66} p. 114
                DOI 10.1238/Physica.Regular.066a00114

        \bibitem{ZhouWang}
                Zhou, Y. and Wang, X. 2022 A methodology for formulating dynamical equations in analytical mechanics based on the principle of energy conservation \textit{J. Phys. Commun.} \textbf{6} p. 035006
                DOI 10.1088/2399-6528/ac57f8

	\bibitem{HancTaylor}
                Hanc, J. and Taylor, E. F. 2004 From conservation of energy to the principle of least action: A story line \textit{Am. J. Phys.} \textbf{72} p. 514
		DOI 10.1119/1.1645282

	\bibitem{NTT}
        	Neuenschwander, D. E., Taylor E. F. and Tujela, S. 2006 Action: forcing energy to predict motion \textit{The Physics Teacher} \textbf{44} p. 146
		DOI 10.1119/1.2173320

	\bibitem{MarionThornton}
		Thornton S. T. and Marion J. B. 2004 \textit{Classical dynamics of particles and systems} 5th Ed (Belmont: Thomson Brooks/Cole) pp. 80-81

        \bibitem{Arnold}
                Arnold V. I. 1989 \textit{Mathematical Methods of Classical Mechanics} 2nd Ed (New York: Springer-Verlag). pp. 16,22

        \bibitem{Taylor}
                Taylor J. R. 2005 \textit{Classical Mechanics} (California: University Science Books) p. 404, 431-434

        \bibitem{Shinbrot}
                Shinbrot, T., Grebogi, C. Wisdom, J. and Yorke, J. A. 1992 Chaos in a double pendulum \textit{Am. J. Phys.} \textbf{60} p. 491

        \bibitem{Vadai}
                Vadai, G., Gingl, Z. and Mellár, J. 2012 Real-time demonstration of the main characteristics of chaos in the motion of a real double pendulum \textit{Eur. J. Phys.} \textbf{33} pp. 907-920
                DOI 10.1088/0143-0807/33/4/907

        \bibitem{Korsch}
                Korsch, H. J. and Jodl, H. J. 1994 \textit{Chaos: a program collection for the PC} (Berlin Heidelberg: Springer) pp. 89-114

        \bibitem{Rafat}
                Rafat, M. Z., Wheatland, M. S. and Bedding, T. R. 2009 Dynamics of a double pendulum with distributed mass \textit{Am. J. Phys.} \textbf{ 77} p. 216
                DOI 10.1119/1.3052072


\end{thebibliography}
\end{document}